\documentclass{ws-procs9x6}
\def\la{\langle}
\def\ra{\rangle}
\def\beq{\begin{equation}}
\def\eeq{\end{equation}}
\def\be{\begin{eqnarray}}
\def\ee{\end{eqnarray}}
\def\hs{\hat{s}}
\def\htm{\hat{t}}
\def\hu{\hat{u}}

\def\k2av{\la k_T^2\ra}

\newcommand{\f}[2]{\frac{#1}{#2}}
\newcommand{\dd}{ {\textrm d}}

\begin{document}

\title{High-$p_T$ Pion Production in Heavy-Ion Collisions at RHIC energies}

\author{G.~G. BARNAF{\"O}LDI, P. L{\'E}VAI}

\address{ KFKI Research Institute for Particle and Nuclear Physics, \\
P.O. Box 49, Budapest, 1525, Hungary \\
E-mail: bgergely@rmki.kfki.hu}

\author{G. PAPP}

\address{HAS Research Group for Theoretical Physics, E\"otv\"os University \\
P{\'a}zm{\'a}ny P{\'e}ter s{\'e}t{\'a}ny 1/A, Budapest, 1117 Hungary}

\author{G. FAI and Y. ZHANG}

\address{ Center for Nuclear Research, Department of Physics,\\
Kent State University, Kent, OH 44242, USA}


\maketitle

\abstracts{Perturbative QCD results on $\pi$ production are presented 
in proton-proton, proton-nucleus and nucleus-nucleus collisions from  
CERN SPS up to RHIC energy. A $K_{jet}\left(s, p_T, Q \right) $ factor 
obtained from jet production is applied to perform next-to-leading order
calculations. Using the intrinsic transverse momentum $(k_T)$ we
determined transverse momentum spectra for pions in wide energy region. 
We have investigated nuclear multiscattering and the Cronin effect at
RHIC energies.}


\section{Introduction}
\label{sec_intro}

The calculation of high-$p_T$ particle production at SPS and RHIC energies 
requires solid theoretical background, based on perturbative QCD (pQCD). 
In these calculations, the choice of renormalization, factorization 
and fragmentation scales are typically chosen in the range 
$p_T/3 \leq Q \leq 2p_T$. Earlier we have found~\cite{BGG} that the 
theoretical reproduction of the measured nuclear effects in 
$pA$ and $AA$ collisions (specifically the Cronin effect~\cite{cronin75}) 
strongly depends on the choice of the above scales. In the following
we display new results analysing preliminary RHIC data~\cite{QM12} 
on pion production in 
$AuAu \rightarrow \pi^0$ at $\sqrt{s} = 130$ AGeV and $200$ AGeV. 


\section{Perturbative QCD calculations with intrinsic $k_T$}
\label{sec_kT}

In pQCD-improved parton model (including intrinsic $k_T$) pion production was
calculated in leading order (LO)~\cite{plf00,zfpbl02} and next-to-leading 
order (NLO)~\cite{pappnlo}. An intermediate solution to speed up the time 
consuming full NLO calculations is to determine a $K_{jet}(s ,p_{T,jet},Q)$
factor corresponds to higher order contributions at jet level in $pp$ 
collisions~\cite{bflpz01} and apply it in the factorized pQCD equation.
In this case the pion production in $AA$ collision can be described as: 
\begin{eqnarray}
 E_{\pi}\f{\dd \sigma_{\pi}^{AA}}{\dd ^3p} &=&
 \sum_{abcd} \,\, \int\! \dd^2 b \, \dd^2 r \,\, t_A(r) \,\, t_B(|\vec b - \vec r|)  \int\! 
\dd ^2 k_{T,a} \, \dd ^2 k_{T,b} \,\dd x_a \, \dd x_b \, \dd z_c \times \ \nonumber \\ 
 \  & \times  & \ g_{pA}({\vec k}_{T,a}) \ f_{a/A}(x_a,Q^2) \,\,  g_{pA}({\vec k}_{T,b}) \ 
f_{b/A}(x_b,Q^2)  \times \ \nonumber \\
 \  & \times  &  \left[ K_{jet}(s ,p_{T,jet},Q)  
 \, \, \f{\dd \sigma}{\dd \htm}^{ab \rightarrow cd}\,
 \right]  \frac{D_{\pi/c}(z_c, Q'^2)}{\pi z_c^2} 
 \,\, \hs \,\, \delta(\hs+\htm+\hu)\ . 
\label{hadX}
\end{eqnarray}

\noindent
The partonic cross section, $\dd \sigma/ \dd\htm$, is the LO Born term, 
which is multiplied by $K_{jet}( s, p_{T,jet},Q)$  
in eq.~(\ref{hadX}). ($\hs$, $\htm$ and $\hu$ are parton-level 
Mandelstam variables.) 
The nuclear thickness function, $t_A(b) = \int \dd z \, \rho(b,z)$, is 
normalized as $\int \dd ^2b \, t_A(b) = A$. The nuclear parton 
distribution functions (PDF) $f_{a/A}(x, Q^2)$ are based on the nucleon 
PDFs ($f_{a/p}(x_a,Q^2)$) and modified in the nuclear environment 
(``shadowing'')~\cite{wang91}. We are using NLO MRST~\cite{MRST} PDF set with 
$Q=p_{T,jet}$ scale, where $p_{T,jet}=p_T/z_c$.

In a phenomenological approach, we introduced an extra 2-dimensional Gaussian
 transverse momentum distribution as the PDF's extension~\cite{zfpbl02}:
\beq
\label{kTgauss}
g_{pA}({\vec k}_T) = \f{1}{\pi \la k^2_T \ra _{pA}}
        \,\, e^{-{k^2_T}/{\la k^2_T \ra}_{pA}} ,  \,\,   
\textrm{where}  \,\, \k2av_{pA} = \k2av_{pp} + C \! \cdot \! h_{pA}(b) .
\eeq
We have found the Gaussian width ($\k2av_{pp}$) to be $p_T$ independent in 
the window $2$ GeV $\leq p_T \leq 6 $ GeV. Nuclear multiscatterings yield
an extra broadening in $pA$ collision, which can be related to the number
of nucleon-nucleon (NN) collisions in the medium and denoted by
$C \! \cdot \! h_{pA}(b)$. Here $C $ means the average increase in the 
transverse momentum width in one NN collision and $h_{pA}(b)$ is a 
geometrical effectivity function. This latter quantity can be 
saturated~\cite{plf00,zfpbl02}. 
The $D_{\pi/c}(z_c, Q'^2)$ is the fragmentation function (FF), which gives 
the probability for parton $c$ to fragment into $\pi$ at momentum fraction 
$z_c$ and fragmentation scale $Q'$. In this paper we fix $Q' = p_T/2$ and 
use the KKP parametrization\cite{KKP}.

 
\section{pQCD result on $pp$ and $AuAu$ at RHIC energies}
\label{sec_cronin}

In Figure 1 we display our results at $\sqrt s =130$ AGeV {\sl (left panels)}
and $200$ AGeV {\sl (right panels)}. In the upper panels the dashed 
curves indicate the $pp$ results which can be compared to the preliminary 
PHENIX data\cite{QM12} {\sl (full triangles)} at $\sqrt s =200$ GeV. 
Full dots indicate the $\pi ^0$ data from peripheral $AuAu$ collisions,
the calculated pion spectra {\sl (solid lines)} approximately overlap with the
data. We obtained $N_{bin,p}^{130}=19 $ binary collisions 
for $60-80 \% $ peripheral case 
and $N_{bin,p}^{200}=10$ for $70-80 \% $ peripheral case. 
A satisfactory agreement between the data and the 
calculations demonstrate the validity of the improved pQCD 
calculation\footnote{Parameters: $ \k2av_{pp}^{130}=1.8$ GeV$^2$, 
$ \k2av_{pp}^{200}=1.5$ GeV$^2$, $C_{sat}=0.4$ and $\nu_{max}=3$.}. 

Full squares indicate the data for central collisions. The number of 
binary collisions are $N_{bin,c}^{130}=857$ for $10 \% $ central $AuAu$ case
at $\sqrt{s}=130$ AGeV, and $N_{bin,c}^{200}=940 $ at
$\sqrt{s}=200$ AGeV. One can find a large 
difference between central data and theoretical spectra {\sl (full lines)}, 
a suppression factor of $\sim 4-5$ appeared. 
\begin{figure}[ht]
\begin{center}
\epsfxsize=3.5in 
\epsfbox{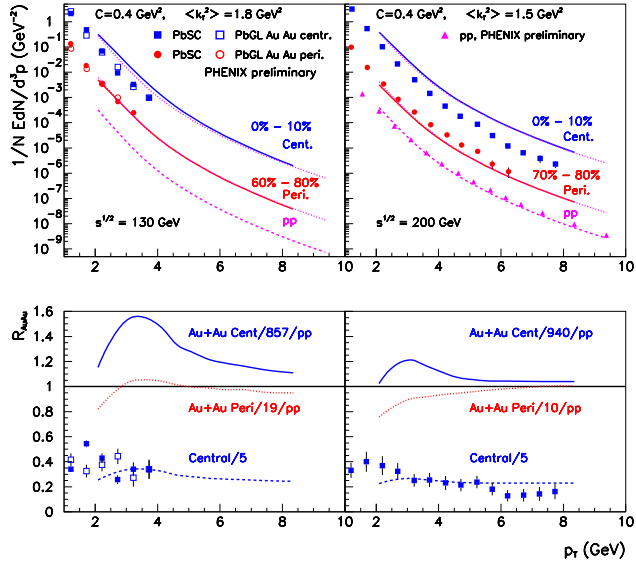}
\caption{ 
$K_{jet}$ based NLO pQCD $\pi^0$ $p_T$-spectra 
{\sl (upper panels)} and the nuclear modification factor $R_{AA}^{ \pi }$
{\sl (lower panels)} compared to preliminary PHENIX $AuAu$ and $pp$
data$^3$ at $\sqrt s = 130$ AGeV {\sl (left)} and $200$ AGeV 
{\sl (right)} RHIC energies. (See details in text.)
}
\label{fig1}
\end{center}
\end{figure}

Lower panels display the nuclear modification factor, defined as
\begin{equation} 
R_{AA}^{\pi}=\frac{1}{\langle N_{bin}\rangle } \frac{\sigma_{in}^{pp}}
{\sigma_{in}^{AA}(b)} \frac{ \dd \sigma_{AA}/ \dd^3 p_{\pi} }
{ \dd \sigma_{pp}/ \dd ^3p_{\pi} } \ .
\end{equation}
This expression
magnifies the enhancement of particle production in $AuAu$, relatively 
to the $N_{bin}$ upscaled $pp$ collision on a linear 
scale. Applying shadowing\cite{wang91} and 
multiscattering, the Cronin peak can be seen clearly in both peripheral 
{\sl (dotted lines)} and central {\sl (solid lines)} collisions. 
The position of Cronin peak depends on the $\k2av_{pp}$ value, 
and the height depends on the extra
broadening\cite{BGG}, $ C \! \cdot \! h_{pA}(b)$. One can see that the 
height of the peak in central case is larger than in peripheral one, 
which effect is caused by the number of nucleon-nucleon collisions 
inside a nucleus. We 
have fixed the PDF scale $Q=p_{T,jet}$ and FF scale $Q'=p_T/2$  
to obtain the peak's maximum at $ p_T \approx 3$ GeV/c.
The comparison of PHENIX data {\sl (full squares)} with theoretical 
calculations display  a difference 
of $\sim 5$. This suppression value is indicated on lower panels  with
a dashed lines. The origin of this suppression is
jet quenching\cite{QM01}.

\newpage 
\section*{Acknowledgments}
\label{sec_ack}

This work was supported in part by  U.S. DOE grant DE-FG02-86ER40251, NSF 
grant INT-0000211, FKFP220/2000, Hungarian grants OTKA T032796, 
T034842, and the HAS $-$ Dubna agreement. Supercomputer time provided by 
BCPL in Norway and the EC $-$ Access to Research Infrastructure action of 
the Improving Human Potential Programme is gratefully acknowledged, together 
with the ITOL at E\"otv\"os University.


\vfill\eject
\end{document}